\documentclass[amsmath,amssymb,amsbsy,prb,twocolumn,preprintnumbers,showpacs,superscriptaddress]{revtex4-2}
\usepackage{graphicx,color}
\usepackage{dcolumn}
\usepackage{bm}
\usepackage{braket}
\usepackage{mathtools}
\usepackage{ulem}
\usepackage{mathrsfs}
\usepackage{times}
\usepackage[breaklinks,colorlinks=true,linkcolor=blue,urlcolor=blue,citecolor=blue]{hyperref}

\newcommand{\Bolz}{k_{\rm B}}

\begin{document}
\title{Thermoelectric transport of type-I, II, and III massless Dirac fermions in two-dimensional lattice model}

\author{Tomonari Mizoguchi}
\affiliation{Department of Physics, University of Tsukuba, 1-1-1 Tennodai, Tsukuba, Ibaraki 305-8571, Japan}
\email{mizoguchi@rhodia.ph.tsukuba.ac.jp}
\author{Hiroyasu Matsuura}
\affiliation{Department of Physics, University of Tokyo, 7-3-1 Hongo, Bunkyo-ku, Tokyo 113-0033, Japan}
\author{Masao Ogata}
\affiliation{Department of Physics, University of Tokyo, 7-3-1 Hongo, Bunkyo-ku, Tokyo 113-0033, Japan}
\affiliation{Trans-scale Quantum Science Institute, University of Tokyo, Bunkyo-ku, Tokyo 113-0033, Japan}

\date{\today}
\begin{abstract}
We study longitudinal electric and thermoelectric transport coefficients of Dirac fermions on a simple lattice model 
where tuning of a single parameter enables us to change the type of Dirac cones from type-I to type-II. 
We pay particular attention to the behavior of the critical situation, i.e., the type-III Dirac cone. 
We find that the transport coefficients of 
the type-III Dirac fermions behave as the limiting case of neither the type-I nor type-II.
On the one hand, the qualitative behaviors of the type-III case are similar to those of the type-I case.
On the other hand, the transport coefficients do not change monotonically upon increasing the tilting; namely,
the largest thermoelectric response is obtained not for the type-III case but for the optimally tilted type-I case.
For the optimal case, the sizable transport coefficients are obtained; 
for example, the dimensionless figure of merit is 0.18.
\end{abstract}

\maketitle
\section{Introduction}
In the past few decades, Dirac fermions in solids have attracted considerable interest.
In particular, two-dimensional systems hosting Dirac cones have been intensively investigated, both theoretically and experimentally.
Graphene, a single-layered honeycomb network of carbon atoms, is a prime example 
of massless Dirac-fermion systems~\cite{Wallace1947,Novoselov2004,CastroNeto2009}.
The organic conductor $\alpha$-(BEDT-TTF)$_2$I$_3$ [BEDT-TTF is bis(ethylenedithio)-tetrathiafulvalene]~\cite{Kajita2014,Katayama2006,Kobayashi2007,Fukuyama2007,Goerbig2008,Kobayashi2008,Kobayashi2009} is another example of the massless Dirac-fermion systems in quasi-two dimensions.
An interesting feature of $\alpha$-(BEDT-TTF)$_2$I$_3$ is that the 
Dirac cones are not isotropic in momentum space, that is, the cones are tilted. 
Triggered by this finding, the effects of tilting of Dirac cones have been investigated. 
It was revealed that Dirac cones are classified into three types according to the degree of tilting:
The tilted Dirac cone with the ellipsoidal equi-energy surface around the Dirac point is classified as type-I.
By further increasing the tilting, the Dirac cones are ``overtilted" and the equi-energy surface turns into a hyperbola;
such a Dirac cone is classified as type-II.
The critical point between type-I and type-II is called a type-III Dirac cone, 
where one of the bands composing the Dirac cone has flat dispersion along a certain direction, resulting in a diverging density of states (DOS) at the Dirac point. 
Although the type-III Dirac cone is rare compared with the other two types
because it does not appear as a stable ``phase" occupying a finite region of the parameter space,
it has gained attention recently~\cite{Volovik2016,Volovik2017,Volovik2018,Liu2018,Huang2018,Fragkos2019,Milicevic2019,Farajollahpour2019,Kim2020,Chen2020,Jin2020,Gong2020,Jin2020_2,Farajollahpour2020,Fragkos2021,Li2021,Sims2021}.
\begin{figure}[b]
\begin{center}
\includegraphics[clip,width = 0.95\linewidth]{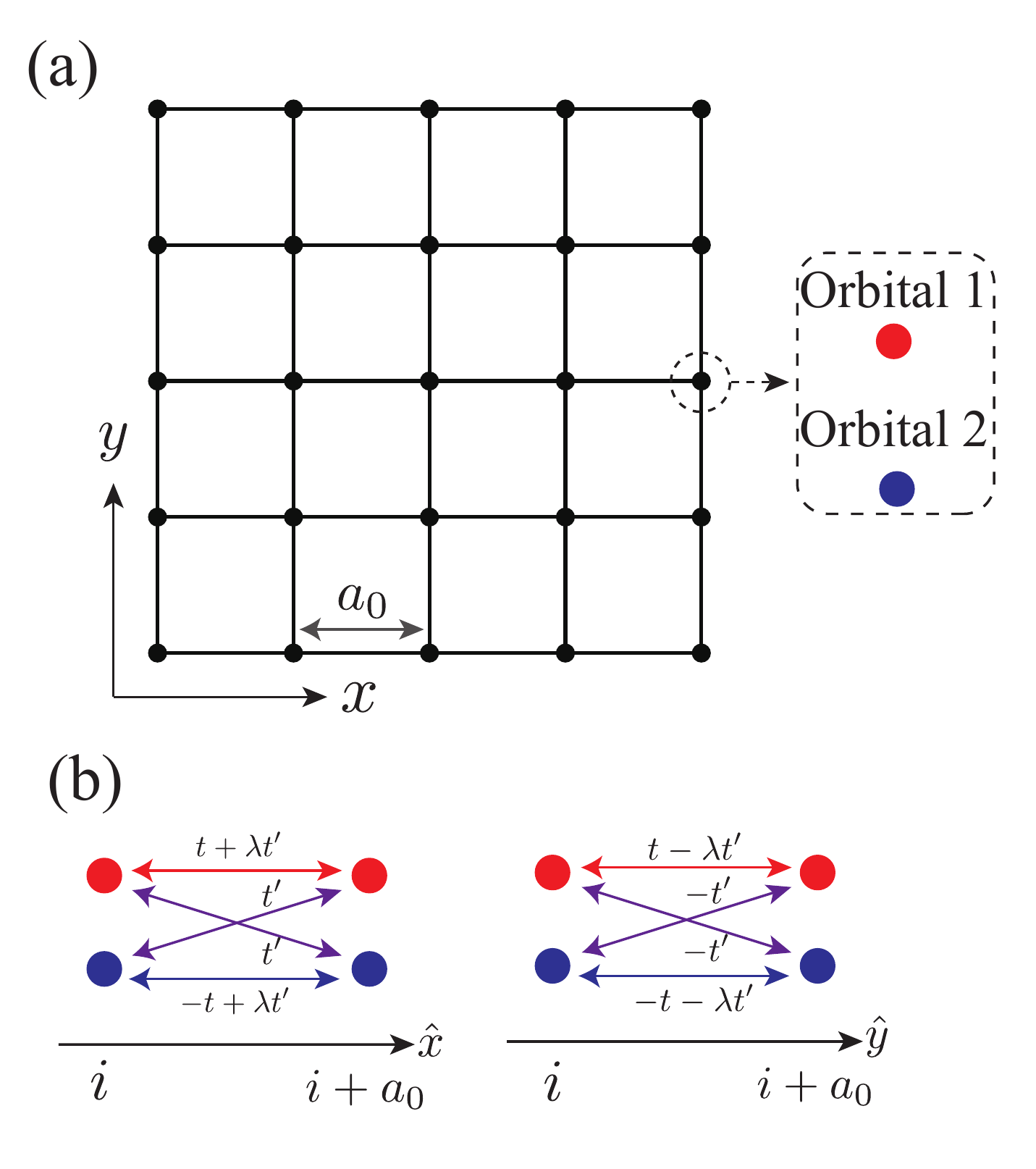}
\vspace{-10pt}
\caption{Schematic figure of the tight-binding model of Eq.~(\ref{eq:Ham_sq}).
(a) The lattice structures and (b) the hopping processes in the $x$ and $y$ directions. The red and blue dots denote orbitals 1 and 2, respectively.}
  \label{fig:model}
 \end{center}
 \vspace{-10pt}
\end{figure}

Along with studies from the viewpoint of electronic structure, 
exotic transport~\cite{Fukuyama2007,Kobayashi2008,Suzumura2014,Suzumura2014_2,Proskurin2015,Mani2017,Mani2019,Mani2019_2} and magnetic properties~\cite{Koshino2007,Koshino2010,GomezSantos2011,Raoux2015,Ogata2016} 
of Dirac fermions have also been studied.
The main targets of such studies are type-I and type-II Dirac cones,
and thus the properties of 
type-III Dirac cones are less understood compared with the other two types. 
Since type-III is the 
critical point between type-I and type-II, one may ask the following question:
Can we understand the behavior of type-III Dirac cones by taking the limit from type-I or type-II?

So far, electric and thermoelectric transports for tilted Dirac fermions, 
both the longitudinal and transverse ones, have been intensively studied~\cite{Ferroiros2017,Mani2017,Kozii2019,Rostamzadeh2019,Mawrie2019,Mandal2020,Ohki2020}.
However, previous works were mostly on continuum models, and research on lattice models is limited.
(For instance, for the three-dimensional case, studies on the minimal lattice model~\cite{McCormick2017_TB} 
were reported in Refs.~\cite{McCormick2017,McKay2019}.) 
In the continuum model, however, there is a subtlety to the momentum cut-off dependence;
namely, for types II and III, the Fermi surface extends far away from the Dirac point, 
where the Dirac-Hamiltonian description is broken down in actual materials.
This hampers the study on the transport coefficients of all the three types in an equal-footing manner within the continuum model,
which motivates us to study a lattice model.

In this paper, we study the longitudinal transport coefficients of a simple lattice model with Dirac cones in two dimensions. 
The model is a generalization of a model proposed by one of the authors~\cite{Mizoguchi2020_typeIII}, 
in which the type-III Dirac cone is realized.
The slight modulation of the Hamiltonian enables us to control the type of Dirac cones by a single parameter, as we will show later.
Therefore, the model serves as a minimal model of tilted Dirac cones in two dimensions. 
\begin{figure*}[t]
\begin{center}
\includegraphics[clip,width = 0.98\linewidth]{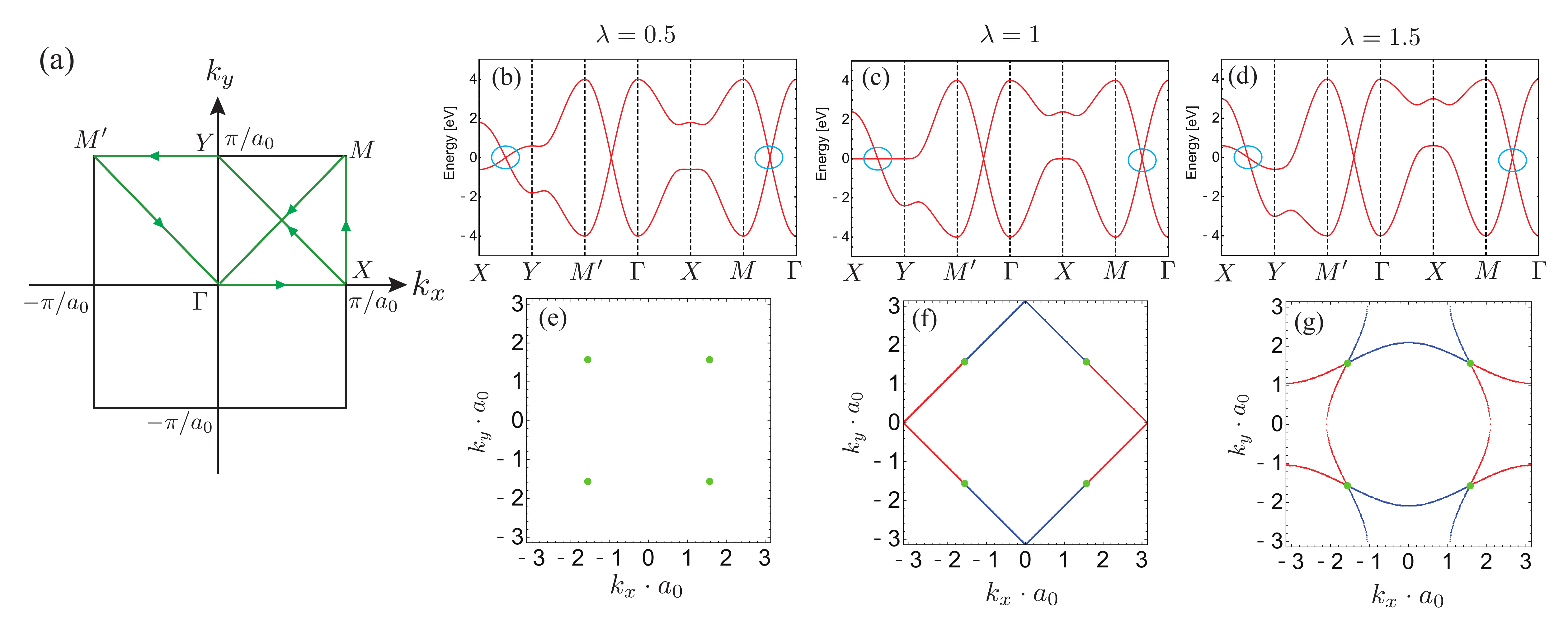}
\vspace{-10pt}
\caption{
(a) The first Brillouin zone. The green lines correspond to 
the high-symmetry lines on which we plot the band structure in (b)-(d).
Band structures for the Hamiltonian in
Eq.~(\ref{eq:Ham_sq}) with $t=-1$ eV, $t^\prime = -0.3$ eV, and
(b) $\lambda = 0.5$, (c) $\lambda = 1$, and (d) $\lambda = 1.5$.
The horizontal axis denotes $\bm{k}$. 
The Dirac cone at $\bm{k} = \left( \frac{\pi}{2 a_0}, \frac{\pi}{2 a_0} \right)$ are encircled by cyan circles. 
The Fermi surface at $\mu = 0$ eV for (e) $\lambda = 0.5$, (f) $\lambda = 1$, and (g) $\lambda = 1.5$.
The green dots represent the Dirac points. Red and blue lines are the electron-type surface (i.e., $\varepsilon_{\bm{k},+} = \mu$) and the 
hole-type surface (i.e., $\varepsilon_{\bm{k},-} = \mu$), respectively.
}
  \label{fig:band}
 \end{center}
 \vspace{-10pt}
\end{figure*}

For this model, we calculate the electric and thermoelectric transport coefficients 
on the basis of the  Kubo formula~\cite{Kubo1957,Luttinger1964}.
We consider the effects of nonmagnetic impurities by using the relaxation time approximation.  
Our result indicates that the largest thermoelectric response is obtained not for the type-III case but for the optimally tilted type-I case.
This indicates that the thermoelectric transport coefficients of type-III Dirac fermions cannot be regarded as 
a limiting case of either type-I or type-II.
To be more specific, the transport coefficients of type-III Dirac fermions are qualitatively similar 
to those of type-I, in that 
the spectral conductivity shows a dip rather than a peak at the Dirac point for the type-I and the type-III cases, 
and that the sign of the Seebeck coefficient for the type-III case is the same as that for type-I case.
However, the transport coefficients do not behave monotonically upon increasing the tilting.
Quantitatively, for the optimal case, sizable transport coefficients are obtained; for example, 
the dimensionless figure of merit is 0.18 for the temperature of the order of 100 K under a trial setting of parameters.

The rest of this paper is structured as follows.
In Sec.~\ref{sec:model}, we introduce the model considered in this paper, namely, 
a square-lattice model with two internal degrees of freedom.
The main results of this paper are presented in Sec.~\ref{sec:results}.
We first show the chemical potential dependence of the electric conductivity at zero temperature. 
Then we argue the temperature dependence of the Seebeck coefficient, the power factor, and the dimensionless figure of merit.
Finally, we present the results in the low temperature region based on the Mott formula,
which are helpful for obtaining a deeper understanding about the comparison among the three types of Dirac cones.
A summary of this paper is presented in Sec.~\ref{sec:summary}.

We remark that, throughout this paper, $\hbar$ represents 
the reduced Planck constant, and $\Bolz$ represents the Boltzmann constant. 

\section{Model: Two-orbital square-lattice model \label{sec:model}}
To comprehensively study transport coefficients of Dirac fermions of 
all three types on a lattice model, 
we introduce a simple tight-binding model defined on a square lattice. 
The model is an extension of one introduced in Ref.~\onlinecite{Mizoguchi2020_typeIII}, 
where the type-III Dirac cones are selectively tailored.
The model considered here is a spinless-fermion model. 
If we incorporate the spin degrees of freedom,
the spectral conductivity of Eq.~(\ref{eq:alpha}) is to be multiplied by 2, 
thus the results in Sec.~\ref{sec:results} will be modified accordingly.
The spinless fermions have two internal degrees of freedom, labeled by 1 and 2, which we will call ``orbitals" henceforth [Fig.~\ref{fig:model}(a)]. 

Our tight-binding Hamiltonian is given as 
\begin{eqnarray}
H =\sum_{\langle i,j \rangle} \sum_{\eta_1,\eta_2 = 1,2} t_{i,j}^{\eta_1,\eta_2 } c^\dagger_{i,\eta_1} c_{j,\eta_2} +(\mathrm{H.c.}),
\end{eqnarray}
where $i$ and $j$ denote the sites and $\langle \cdot , \cdot \rangle$ denotes the nearest-neighbor pair of sites.
The hopping integrals $t_{i,j}^{\eta_1,\eta_2 }$ are depicted in Fig.~\ref{fig:model}(b).
We note that the hopping integrals in the $x$ direction are different from those in the $y$ direction.
The momentum space representation is given as
\begin{eqnarray}
H = \sum_{\bm{k}} \bm{\psi}^\dagger_{{\bm{k}}} \mathcal{H} ({\bm{k}})\bm{\psi}_{\bm{\bm{k}}},  \label{eq:Ham}
\end{eqnarray}
where $\bm{\psi}_{\bm{\bm{k}}} = \left(c_{\bm{k},1}, c_{\bm{k},2} \right)^{\rm T}$ 
denotes the annihilation operators of fermions with crystal momentum $\bm{k}$ and 
\begin{eqnarray}
\mathcal{H} ({\bm{k}}) = 
\begin{pmatrix}
\lambda a_{\bm{k}}+ d_{\bm{k}}& a_{\bm{k}} \\
a_{\bm{k}}& \lambda a_{\bm{k}} - d_{\bm{k}} \\
\end{pmatrix}. \label{eq:Ham_sq}
\end{eqnarray}
Here we have introduced 
$a_{\bm{k}} := 2t^\prime \left( \cos k_x a_0 - \cos k_y a_0\right)$
and $d_{\bm{k}} := 2t\left( \cos k_x a_0+ \cos  k_y a_0\right)$,
with $a_0$ being the lattice constant. The dimensionless parameter $\lambda$ is real and nonnegative;
the modification from $a_{\bm{k}}$ to $\lambda a_{\bm{k}}$ in the diagonal matrix elements 
is an extension compared with the previous work~\cite{Mizoguchi2020_typeIII}.
\begin{figure}[b]
\begin{center}
\includegraphics[clip,width = 0.95\linewidth]{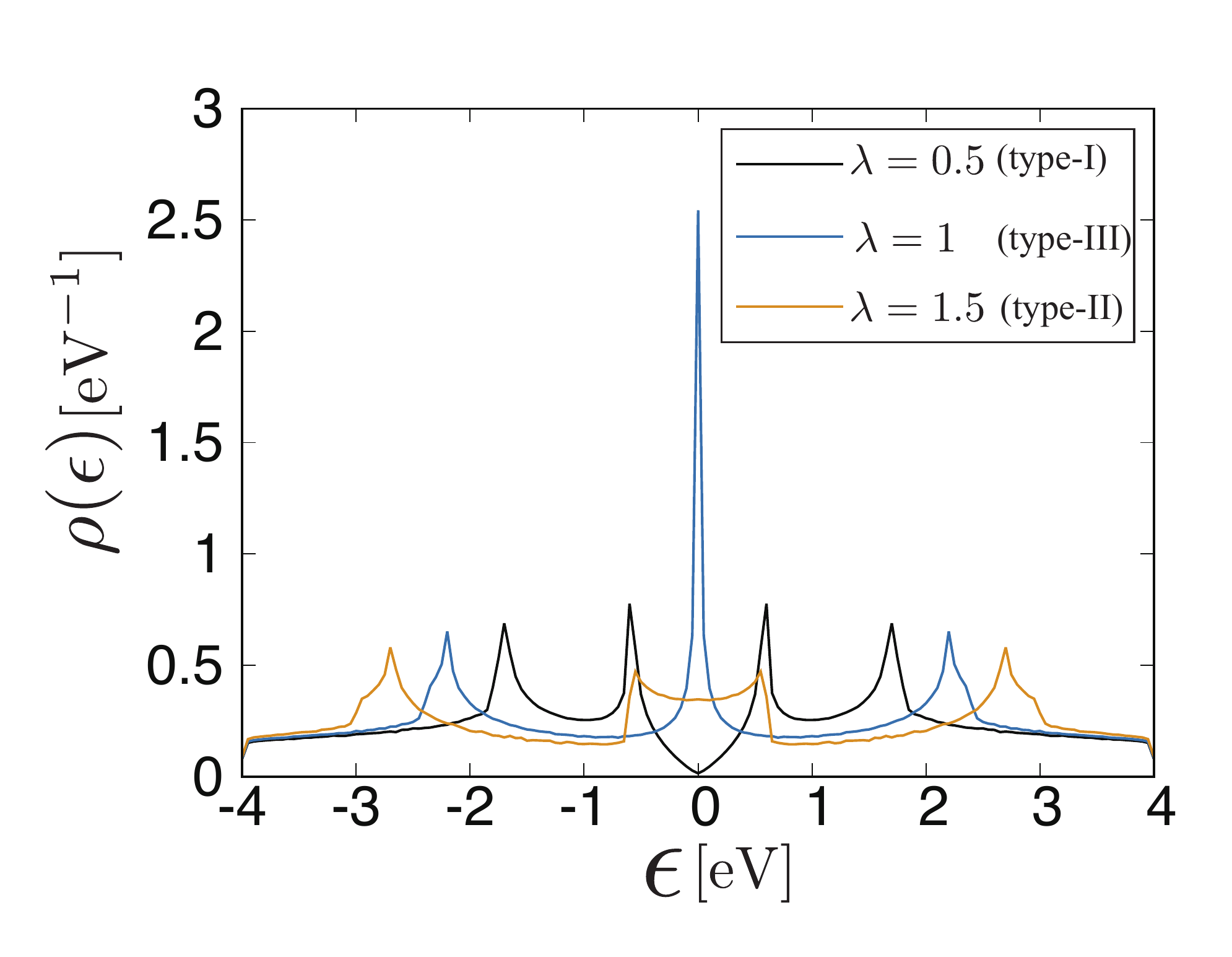}
\vspace{-10pt}
\caption{Density of states for the model in Eq.~(\ref{eq:Ham_sq}).}
  \label{fig:dos}
 \end{center}
 \vspace{-10pt}
\end{figure}

For this model, the dispersion relations of two bands, 
$\varepsilon_{\bm{k}, \pm}$, become
\begin{eqnarray}
\varepsilon_{\bm{k}, \pm} = \lambda a_{\bm{k}}  \pm \sqrt{a^2_{\bm{k}} + d_{\bm{k}}^2 }.  \label{eq:disp}
\end{eqnarray}
In Figs.~\ref{fig:band}(b)-(d), we plot the dispersion relation of Eq.~(\ref{eq:disp}).
Although the model is a toy model and thus the results will not apply to specific materials directly,
it will be useful to set actual values of parameters so that we can roughly estimate the electric and thermoelectric coefficients.
Therefore, in the rest of this paper, we set $t  = -1$ eV and $t^\prime  = -0.3$ eV.
We note that, if the hopping amplitude changes, then the other parameters $\mu$, $\Gamma$, and $T$ should be scaled accordingly.

We find from Eq.~(\ref{eq:disp}) that, for any $\lambda$, 
the Dirac cones appear at the momenta where $a_{\bm{k}} = 0$ and $d_{\bm{k}} = 0$ are simultaneously satisfied,
that is, $\bm{k} = \left(\pm\frac{\pi}{2a_0}, \pm\frac{\pi}{2a_0}\right)$ and $\bm{k} = \left(\pm\frac{\pi}{2a_0}, \mp \frac{\pi}{2a_0}\right)$.
Importantly, the type of the Dirac cone can be tuned by a single parameter $\lambda$, as shown in Fig.~\ref{fig:band}.
Clearly, we have the type-I (type-II) Dirac cones for $\lambda < 1$ ($\lambda > 1$); 
$\lambda = 1$ is the critical case, i.e., the type-III Dirac cones, as pointed out in Ref.~\onlinecite{Mizoguchi2020_typeIII}.
This enables us to study the transport coefficients of three types of Dirac cones comprehensively on this lattice model. 

To further clarify the difference among the thee types, 
we depict the shape of the Fermi surface for $\mu = 0$ eV in Figs.~\ref{fig:band}(e)-(g).
For $\lambda = 0.5$, i.e., for the tilted type-I Dirac cone, the Fermi surface corresponds to the Dirac points.
For $\lambda = 1.5$, i.e., for the type-II Dirac cone, the Fermi surface has a finite area in the Brillouin zone, [Fig.~\ref{fig:band}(g)]
and it consists of two species of surfaces, namely, 
the electron-type surface and hole-type surface, which meet each other at the Dirac points.
For $\lambda = 1$ i.e., for the type-III Dirac cone [Fig.~\ref{fig:band}(f)], 
the Fermi surface shrinks compared with that in Fig.~\ref{fig:band}(g) and forms
straight lines ($k_x \pm k_y = \pi/a_0$, $- \pi/a_0$).

In Fig.~\ref{fig:dos}, we plot the DOS defined as
\begin{eqnarray}
\rho(\epsilon) = -\frac{1}{\pi} \sum_{\bm{k} } 
\mathrm{Im} \left[\frac{1}{\epsilon + i\eta - \varepsilon_{\bm{k},+} } 
+ \frac{1}{\epsilon + i\eta - \varepsilon_{\bm{k},-} } \right],
\end{eqnarray}
where $\eta$ is a small parameter, being set to $0.01|t|$.
For $\lambda = 0.5$, i.e., for the tilted type-I Dirac cone, the DOS drops at $\epsilon = 0$,
reflecting the fact that the Fermi surface consists of the Dirac points.
For $\lambda = 1.5$, i.e., for the type-II Dirac cone, the DOS becomes finite at $\epsilon = 0$,
since the Fermi surface is no longer the points.
For $\lambda = 1$ i.e., for the type-III Dirac cone, a sharp peak of the DOS at $\epsilon = 0$ appears, due to a directionally-flat dispersion at zero energy. 
Away from $\epsilon \sim 0$, we see several peaks for all cases, e.g., $\epsilon \sim$ 0.5 and 2 eV for $\lambda = 0.5$.
They originate from the quasi-flat dispersion near the $X$ and $Y$ points, as shown in Fig.~\ref{fig:band}(b).

\section{Results \label{sec:results}}
\subsection{Longitudinal electric conductivity \label{sec:cond}}
\begin{figure*}[tb]
\begin{center}
\includegraphics[clip,width = 0.95\linewidth]{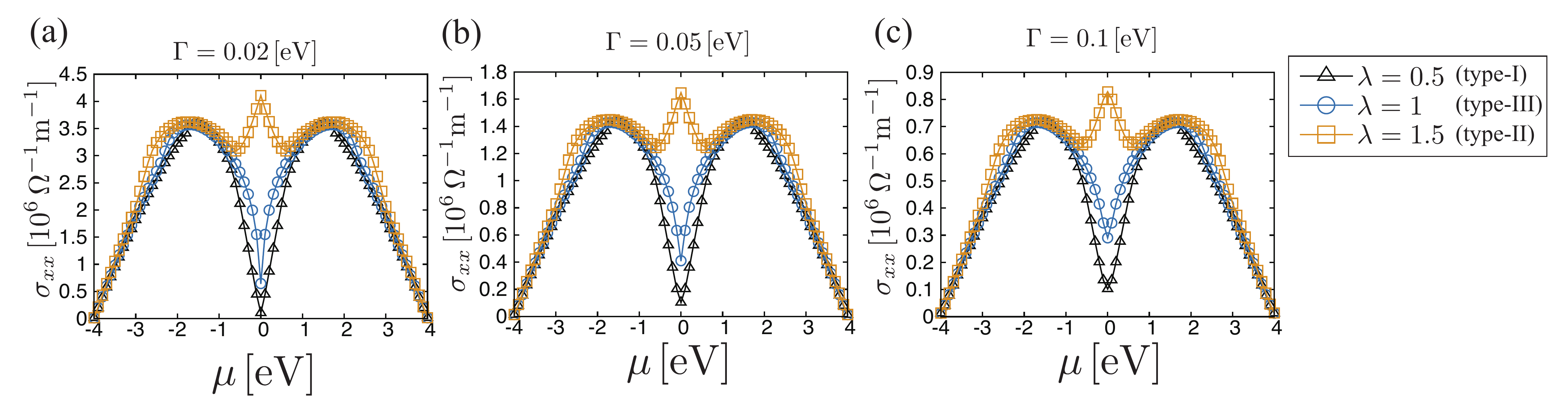}
\vspace{-10pt}
\caption{Longitudinal conductivity for (a) $\Gamma = 0.02$ eV, (b) $\Gamma = 0.05$ eV, and (c) $\Gamma=0.1$ eV.}
  \label{fig:cond}
 \end{center}
 \vspace{-10pt}
\end{figure*}
We first calculate the electric longitudinal conductivity.
The conductivity tensor $\overleftrightarrow{\sigma}$ is defined as 
\begin{eqnarray}
\bm{j}= \overleftrightarrow{\sigma} \bm{E},
\end{eqnarray}
where $\bm{j}$ is the current density and $\bm{E}$ is the electric field. 
The longitudinal conductivity corresponds to the diagonal element of the conductivity tensor, 
which we write $\sigma_{ii}$ ($i=x,y$).
We have confirmed that relation the $\sigma_{xx} = \sigma_{yy}$ holds 
(see Appendix~\ref{app:xy} for the proof), so we focus on $\sigma_{xx}$ henceforth. 
Note that the above relation implies that the anisotropy of the conductivity is not observed in this model,
unlike the case of the continuum model with a single tilted Dirac cone~\cite{Suzumura2014,Suzumura2014_2,Proskurin2015,Rostamzadeh2019}.
This might originate from the fact that there are two pairs of Dirac cones, namely, 
$\bm{k} = \left(\pm\frac{\pi}{2a_0}, \pm\frac{\pi}{2a_0}\right)$ and $\bm{k} = \left(\pm\frac{\pi}{2a_0}, \mp \frac{\pi}{2a_0}\right)$,
whose tilting direction are perpendicular to each other.

The longitudinal conductivity can be calculated by using the Kubo formula:
\begin{eqnarray}
\sigma_{xx} = -  \int_{-\infty}^{\infty} d\epsilon \hspace{0.5mm} f^{\prime}(\epsilon-\mu) \alpha_{xx}(\epsilon),
\label{eq:cond}
\end{eqnarray}
where $\mu$ is the chemical potential and $\alpha_{xx}(\epsilon)$ is referred to as the spectral conductivity:
\begin{eqnarray}
\alpha_{xx}(\epsilon) &=& \frac{\hbar e^2}{2\pi A d_0 } \sum_{\bm{k}}\mathrm{Tr} 
\{ G^{(R)}(\bm{k},\epsilon) v_x (\bm{k}) G^{(A)}(\bm{k},\epsilon) v_x (\bm{k}) \nonumber \\
&-&\mathrm{Re}  \left[G^{(R)}(\bm{k},\epsilon) v_x (\bm{k}) G^{(R)}(\bm{k},\epsilon) v_x (\bm{k})\right] \}.
\label{eq:alpha}
\end{eqnarray}
Here, $-e$ is the charge of an electron, $A$ is the area of the two-dimensional layer,
$f(\epsilon) = 1/(e^{\beta \epsilon} + 1)$ is the Fermi-Dirac distribution function ($\beta = 1/\Bolz T$), and $f^\prime(\epsilon)$ is its derivative.
In Eq.~(\ref{eq:alpha}), we have included an interlayer distance $d_0$ to make $\alpha_{xx}(\epsilon)$ 
have the units of the three-dimensional (bulk) conductivity. 
This means that, although the tight-binding model in Eq.~(\ref{eq:Ham}) is two-dimensional,
we consider the quasi-two-dimensional (quasi-2D) system 
where the independent two-dimensional layers are stacked with interlayer distance $d_0$.
Such a quasi-2D nature applies to
quasi-2D organic materials in which the tilted Dirac electrons are realized and the electric and 
thermoelectric transport coefficients are measured for bulk (three-dimensional) samples.
Hereafter, we set $d_0 = 10 \AA$ as a typical value of the quasi-2D materials.
Note that we have neglected the interlayer coupling.
In general, depending on the symmetries, the Dirac cones can acquire a mass gap 
due to the interlayer coupling. 
Nevertheless, regardless of the existence of such a mass gap, 
a small interlayer coupling does not change the following results qualitatively 
for temperatures greater than the interlayer coupling.

As for the retarded and advanced Green's functions, 
$G^{(R)}(\bm{k},\epsilon)$ and $G^{(A)}(\bm{k},\epsilon)$, respectively, 
we employ the relaxation time approximation:
\begin{eqnarray}
G^{(R)}(\bm{k},\epsilon) = \left[\epsilon + i \Gamma- \mathcal{H}({\bm{k}}) \right]^{-1},
\end{eqnarray}
and
\begin{eqnarray}
G^{(A)}(\bm{k},\epsilon) = \left[ \epsilon - i \Gamma - \mathcal{H}({\bm{k}})\right]^{-1},
\end{eqnarray}
with $\Gamma $ being the damping rate caused by the impurity scattering; we do not consider the screening effect of the impurity potential 
that causes the momentum and frequency dependence of $\Gamma$~\cite{Hwang2007,Hwang2009}. 
We consider three cases, namely, $\Gamma = 0.02$, $0.05$, and $0.1$ eV. 
The velocity $v_x (\bm{k})$ is given as 
\begin{eqnarray}
v_x (\bm{k}) = \frac{1}{\hbar} \frac{\partial  \mathcal{H}(\bm{k})}{\partial k_x}.
\end{eqnarray}

For the numerical calculations, we set $T=0$, where 
$f^{\prime}(\epsilon) =- \delta(\epsilon)$ thus $\sigma_{xx} = \alpha(\mu)$.
Then we numerically take the summation over 
$\bm{k}$ with $800 \times 800$ meshes for $\Gamma = 0.02$ eV,
$400 \times 400$ meshes for $\Gamma = 0.05$ eV, 
and $200 \times 200$ meshes for $\Gamma = 0.1$ eV.

In Fig.~\ref{fig:cond}(a)-(c), we show the $\mu$ dependence of $\sigma_{xx}$. 
The three panels are for different values of $\Gamma$. 
Although $\Gamma$ affects $\sigma_{xx}$  quantitatively, the overall features of $\mu$-dependence 
do not change.
We see that the conductivity sharply drops for $\mu \rightarrow 0$ eV for $\lambda =0.5$ (the type-I case)
and $\lambda =1$ (the type-III case),
while it has a peak at $\mu = 0$ eV for $\lambda = 1.5$ (the type-II case).
In this sense, the conductivity for the type-III case is similar to that for the type-I case, 
rather than the type-II case.

To account for this result, 
we compare the $\mu$ dependence of $\sigma_{xx}$ 
with the DOS profile in Fig.~\ref{fig:dos}.
In general, the finite DOS is essential to obtain a sizable conductivity. 
In this respect, the results for the types-I and II coincide with the DOS profile;
namely, the DOS approaches to zero (finite) at $\mu = 0$ eV for the type-I (II), which is reflected in the $\mu$ dependence of $\sigma_{xx}$.
In contrast, for the type-III case, the conductivity shows a dip at $\mu = 0$ eV despite the peak in the DOS.
This means that the directionally flat dispersion of the type-III Dirac cone, which leads to the peak of the DOS,
does not contribute the conductivity, probably because of the momentum dependence of the velocity operator, 
which is another key factor for the determination of the conductivity.
Such a subtle interplay between the DOS profile 
and the momentum dependence of the velocity operator 
may also lead to the peak at $\mu = 0$ eV for the type-II Dirac cone.

\subsection{Thermoelectric transport coefficients \label{sec:seebeck}}
\begin{figure*}[tb]
\begin{center}
\includegraphics[clip,width =0.99\linewidth]{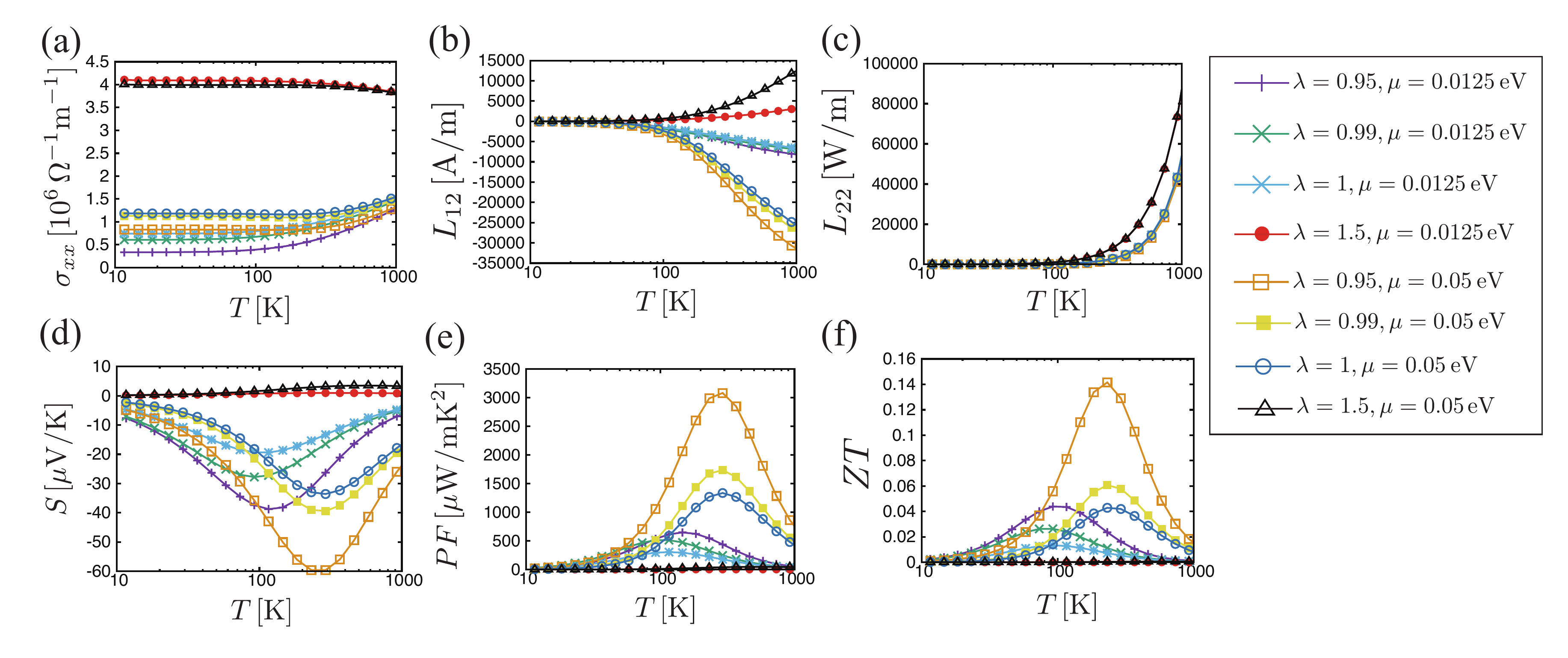}
\vspace{-10pt}
\caption{Temperature dependence of 
(a) the conductivity, (b) $L_{12}$, (c) $L_{22}$,
(d) the Seebeck coefficient, (e) the power factor, and (f) the dimensionless figure of merit.
We set $\Gamma = 0.02$ eV.}
  \label{fig:finite}
 \end{center}
 \vspace{-10pt}
\end{figure*}
Next, we calculate the Seebeck coefficient,~\cite{Luttinger1964,Mahan1990,Yamamoto2018,Ogata2019},
the power factor, and the dimensionless figure of merit.
The definitions of these quantities are as follows.
We focus on the case where the electric and thermal currents as well as the electric field and the temperature gradients
are all in the $x$ direction.
The electric current in the presence of the electric field and the temperature gradient is given as
\begin{eqnarray}
j_x = L_{11} E_x + L_{12} \left(- \frac{\partial_x T}{T} \right),
\end{eqnarray}
and the thermal current {is given as}
\begin{eqnarray}
j^x_Q = L_{21} E_x + L_{22} \left(- \frac{\partial_x T}{T} \right),
\label{eq:JQ}
\end{eqnarray} 
with $\partial_x T$ being the temperature gradient in the $x$ direction. 
Note that $L_{11}=\sigma_{xx}$, and $L_{12}=L_{21}$ due to the Onsager's relation.
The Seebeck coefficient ($S$) is expressed by using $L_{ij}$ ($i,j=1,2$) as
\begin{eqnarray}
S = \frac{1}{T}\frac{L_{12}}{L_{11}}. \label{eq:Seebeck}
\end{eqnarray}
The power factor (PF) and the dimensionless figure of merit ($ZT$) are defined as
\begin{eqnarray}
\mathrm{PF} = \frac{1}{T^2} \frac{L_{12}^2}{L_{11}},
\end{eqnarray}
and 
\begin{eqnarray}
Z T =  \frac{S^2 \sigma}{\kappa} T,
\end{eqnarray}
respectively, where $\kappa$ is the thermal conductivity. 
Generally, $\kappa$ is given by $\kappa=\kappa_e+\kappa_{\rm ph}$, 
and $\kappa_e$ ($\kappa_{\rm ph}$) is the electronic (phonon) contribution to thermal conductivity. 
In this paper, we ignore $\kappa_{\rm ph}$ and calculate $ZT$ using $\kappa_e$ given by
\begin{eqnarray}
\kappa_e = \frac{L_{22}-(L_{12}L_{21})/L_{11}}{T}.
\end{eqnarray}
Thus, the obtained results for $ZT$ are the maximum of the possible $ZT$.
Note that the validity of neglecting the phonon contribution depends on the actual materials.
For instance, in graphene, the phonon contribution is dominant~\cite{JACIMOVSKI2015}.

In the present model, only the impurity scattering potentials 
cause the damping rate, $\Gamma$. 
Therefore, $L_{12}$ is given by~\cite{Jonson1980,Kontani2003,Ogata2019}
\begin{eqnarray}
L_{12} = \frac{1}{e}  \int_{-\infty}^{\infty} d\epsilon \hspace{0.5mm} (\epsilon - \mu) f^{\prime}(\epsilon- \mu) \alpha_{xx}(\epsilon), \label{eq:L12}
\end{eqnarray}
with $\alpha_{xx}(\epsilon)$ being defined as in Eq.~(\ref{eq:alpha}).
This relation between $L_{11}$ and $L_{12}$ is called
the Sommerfeld-Bethe relation~\cite{Sommerfeld1933}.
It should be noted that we consider only the electric contribution to the Seebeck coefficient and 
neglect the other contributions 
such as the phonon drag deriving from the electron-phonon interaction~\cite{Ogata2019,Matsuura2019}. 
Similarly, $L_{22}$ is calculated as
\begin{eqnarray}
L_{22} = - \frac{1}{e^2}  \int_{-\infty}^{\infty} d\epsilon \hspace{0.5mm} (\epsilon - \mu)^2 
f^{\prime}(\epsilon- \mu) \alpha_{xx}(\epsilon). \label{eq:L22}
\end{eqnarray}
\begin{figure*}[tb]
\begin{center}
\includegraphics[clip,width =0.99\linewidth]{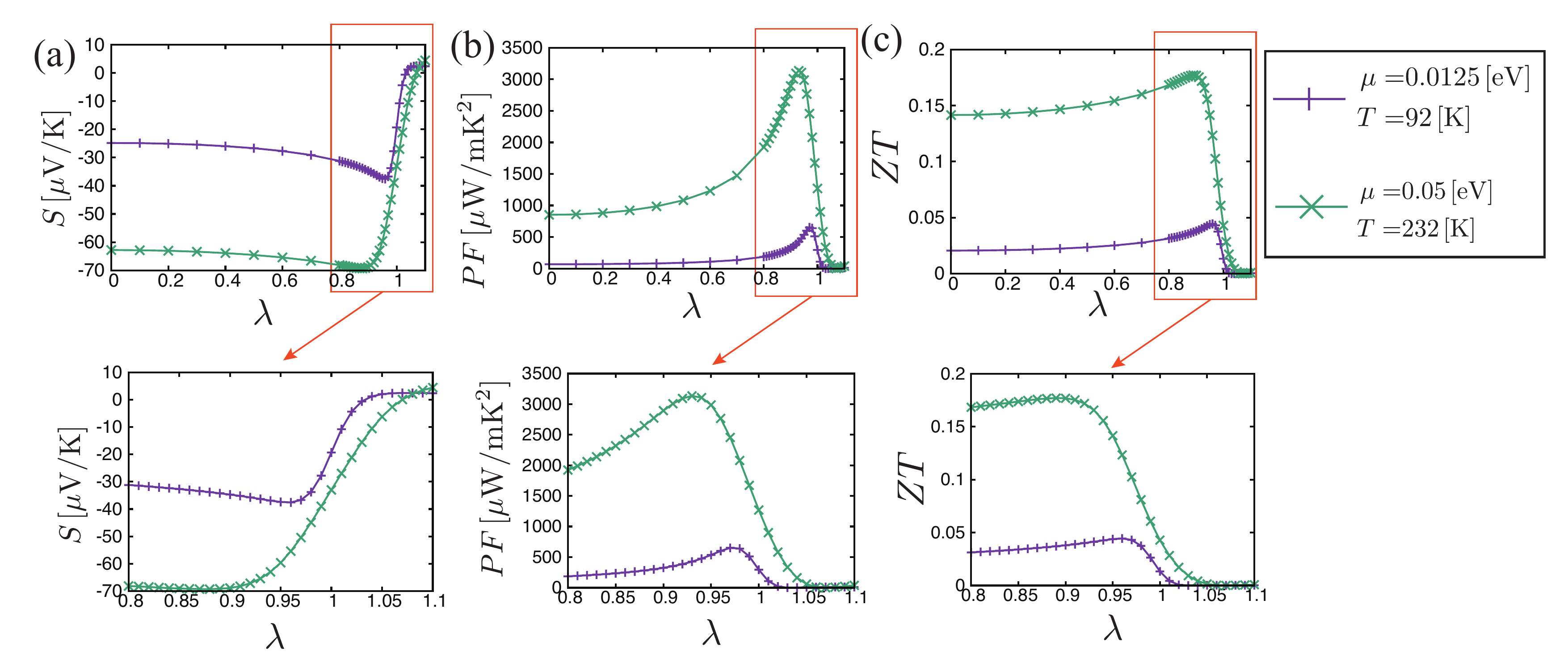}
\vspace{-10pt}
\caption{$\lambda$ dependence of (a) the Seebeck coefficient, (b) the power factor, and (c) the dimensionless figure of merit.
We set $\Gamma = 0.02$ eV.}
  \label{fig:finite_L}
 \end{center}
 \vspace{-10pt}
\end{figure*}
We perform $\epsilon$ integration in Eqs.~(\ref{eq:cond}), (\ref{eq:L12}), and (\ref{eq:L22}) numerically, and calculate 
$S$, PF and $ZT$ by using $L_{11}$, $L_{12}$, and $L_{22}$ thus obtained. 
In the actual numerical calculation, 
we limit the interval of the integration to $\epsilon \in [-1,1]$ (eV) in Eqs.~(\ref{eq:cond}), (\ref{eq:L12}), and (\ref{eq:L22}),
and perform the integration numerically with the number of meshes of $\epsilon$ being 960.
In this section, the damping rate $\Gamma$ is set to be $0.02$ eV.

The results are shown in Fig.~\ref{fig:finite}(a)-(f). 
The eight lines are for the different combinations of $\lambda$ and $\mu$.
Note that we do not consider the temperature dependence of $\mu$ but we set it as a parameter.  
We see that for $\lambda \sim 1$, all three quantities have peaks at temperatures on the order of 100 K.
In fact, for $\lambda \leq 1$, $S$, PF, and $ZT$ have a maximum at
$T \sim \mu/2\Bolz$.
For $S$ and $ZT$, this fact can be accounted for 
by the linear dependence of $\alpha_{xx}(\epsilon)$ as a function of $\epsilon$;
see Appendix~\ref{app:analytic} for details.

Among the combinations of $\lambda$ and $\mu$ shown in Fig.~\ref{fig:finite}(d)-(f), 
the case with $\lambda = 0.95$ and $\mu = 0.05$ eV exhibits the largest response.
The results indicate that there exists an optimal degree of tilting and carrier density to obtain large thermoelectric responses.
We also see that, in deep inside the type-II case ($\lambda = 1.5$), 
the thermoelectric response functions are small compared with those for $\lambda \sim 1$.
Note that, for the type-II case, a dominant contribution to the Seebeck coefficient 
comes from the region near the Dirac points; see Appendix~\ref{app:typeII} for further details.
\begin{figure*}[tb]
\begin{center}
\includegraphics[clip,width = 0.95\linewidth]{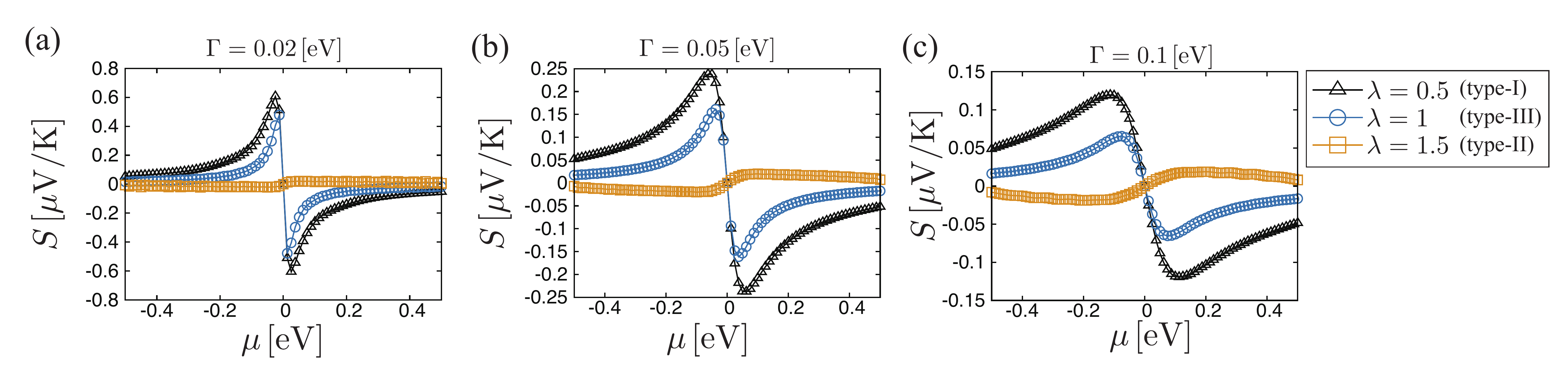}
\vspace{-10pt}
\caption{Seebeck coefficient at $T = 1.16$K for (a) $\Gamma = 0.02$ eV, (b) $\Gamma = 0.05$ eV, and (c) $\Gamma=0.1$ eV.}
  \label{fig:sb}
 \end{center}
 \vspace{-10pt}
\end{figure*}
\begin{figure}[b]
\begin{center}
\includegraphics[clip,width = 0.95\linewidth]{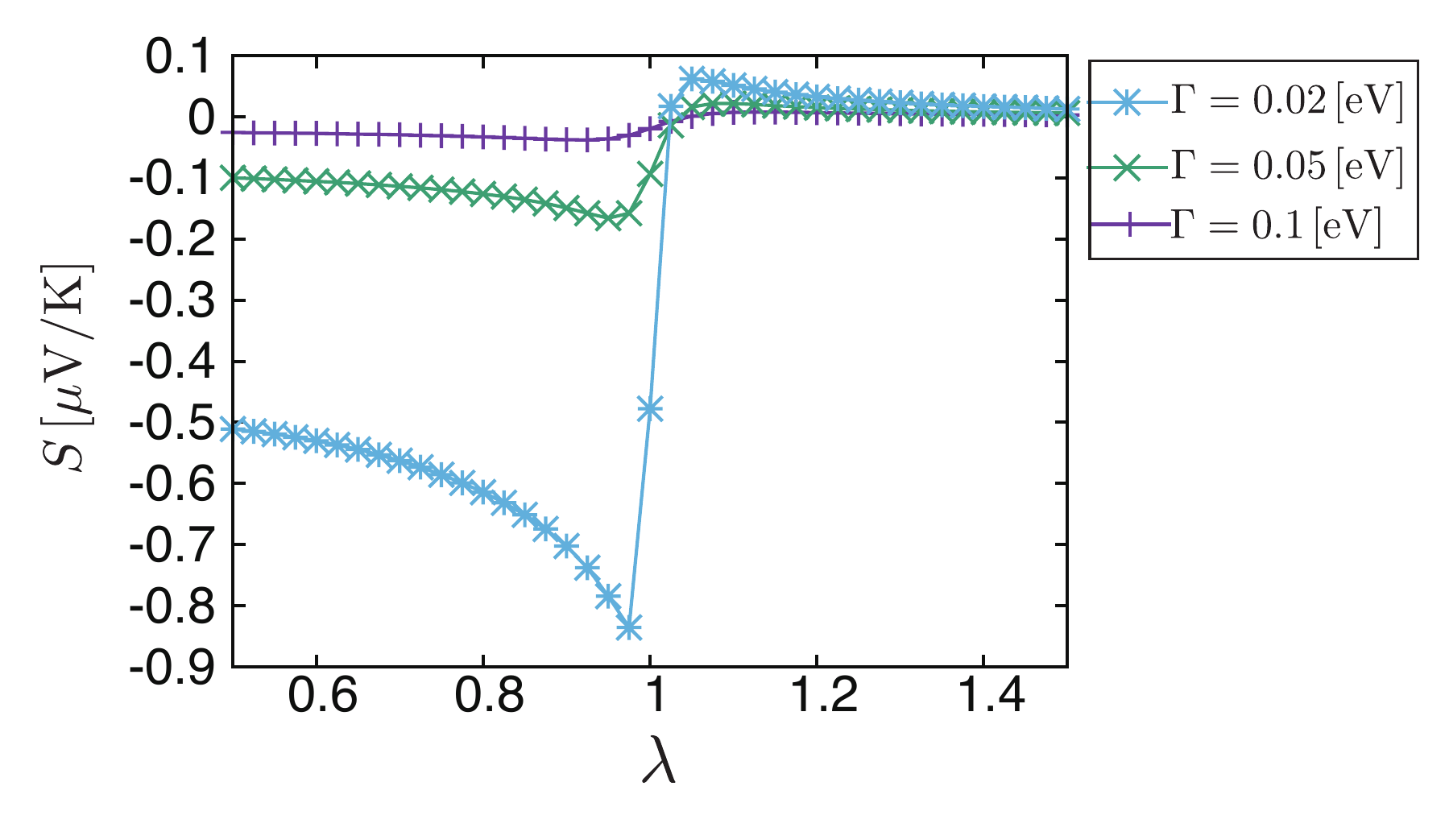}
\vspace{-10pt}
\caption{$\lambda$ dependence of the Seebeck coefficient at $T = 1.16$K and $\mu = 0.0125$ eV.}
  \label{fig:sb_l}
 \end{center}
 \vspace{-10pt}
\end{figure}
\begin{figure*}[tb]
\begin{center}
\includegraphics[clip,width = 0.95\linewidth]{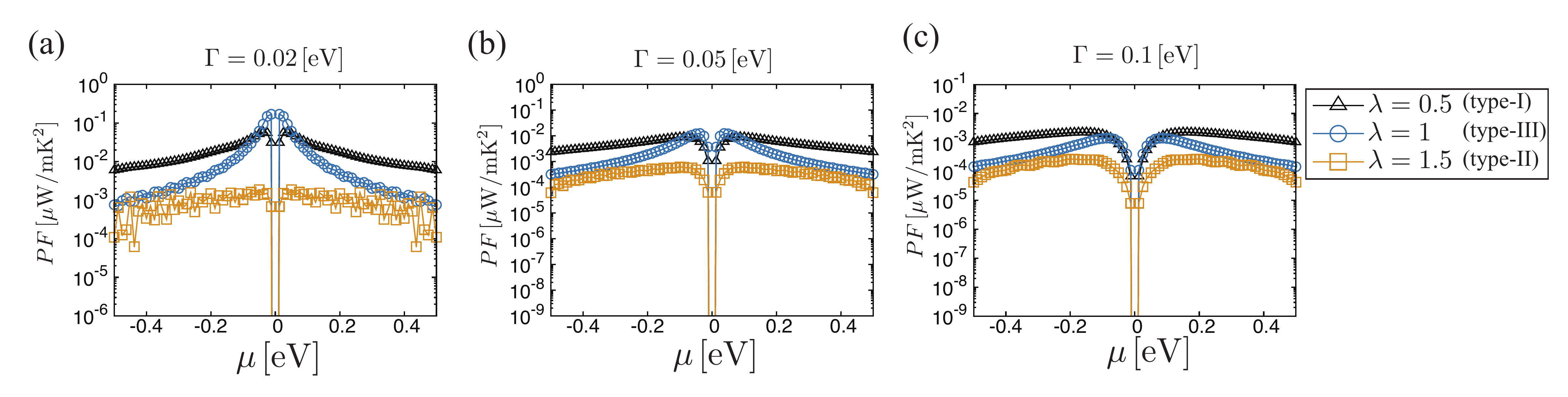}
\vspace{-10pt}
\caption{$\mu$ dependence of the power factor at $T=1.16$ K for  
(a) $\Gamma = 0.02$ eV, (b) $\Gamma = 0.05$ eV, and (c) $\Gamma=0.1$ eV.}
  \label{fig:pf}
 \end{center}
 \vspace{-10pt}
\end{figure*}

To further study the optimal tilting for the large thermoelectric response, 
we investigate the $\lambda$ dependence of $S$, PF, and $ZT$.
The results are shown in Fig.~\ref{fig:finite_L}(a)-(c).
The temperatures are set to be $T = 92$ K and $232$ K for $\mu = 0.0125$ eV and 0.05 eV, respectively,
where the peaks are realized in Fig.~\ref{fig:finite}(d)-(f).
We see that the optimal tilting parameter $\lambda$ indeed exists, and is slightly smaller than $\lambda  = 1$. 
We also see that, for all of the three quantities, their absolute values for $\mu = 0.05$ eV are larger than those for $\mu = 0.0125$ eV.
However, as we will argue in the next section, they do not increase monotonically as a function of $\mu$.
Rather, an optimal value of $\mu$ also exists, as we will explain in the next section. 
Therefore, for the large thermoelectric response, 
the suitable electronic structure is the type-I Dirac cone which is very close to the type-III Dirac cone.
The carrier density (or the chemical potential) also should also be tuned at the optimal value. 
At the maximum within the present results ($\mu = 0.05$ eV and $\lambda = 0.89$), 
we obtain $S \sim -70\hspace{0.5mm}\mathrm{\mu V}/\mathrm{K}$,  
PF $\sim  3000 \hspace{0.5mm}\mu \mathrm{W}/\mathrm{mK}^2$,
and $ZT \sim 0.18$, which are sizable values. 

\subsection{Low temperatures \label{sec:LowT}}
To understand the physical origin of the large Seebeck coefficient and large $ZT$, it is useful to study their low-temperature behaviors.
To this end, for $L_{12}$, we apply the Sommerfeld expansion to Eq.~(\ref{eq:L12}).
Then, we find that, for low temperatures, $S$ is given by the Mott formula~\cite{Mott1936},
\begin{eqnarray}
S = -\frac{\pi^2}{3} \frac{\Bolz^2 T}{e} \left( \frac{d \ln \alpha_{xx}(\epsilon) }{d \epsilon} \right)_{\epsilon= \mu}. \label{eq:L12_lowT}
\end{eqnarray}

In Fig.~\ref{fig:sb}(a)-(c), we show the $\mu$ dependence of $S$ obtained from the Mott formula for $\mu \in [-0.5,0.5]$ eV.
We set the temperature to be small but finite, as $T= 1.16$ K 
(i.e., $\Bolz T = 10^{-4} |t|$).
We see that $S$ is vanishing when 
$\mu$ is right at the Dirac point for all the three types, as 
$S$ is the odd function of $\mu$.
Comparing the three types, 
we find that the type-I and type-III cases have large $S$, while the type-II case has small $S$.
This can be accounted for by the fact that the conductivity (or $L_{11}$) is large for the type-II case.

We also find that the sign of $S$ in the type-I and type-III cases is opposite that for the type II case. 
For instance, for positive $\mu$, $S$ is negative for $\lambda = 0.5$ and $1$, while 
$S$ is positive for $\lambda = 1.5$.
To further clarify the nature of the sign change, 
we plot the $\lambda$ dependence of $S$ at $\mu =0.0125$ eV in Fig.~\ref{fig:sb_l}. 
As $\lambda$ approaches $1$ from the type-I region, $S$ is negative, and $|S|$ becomes larger.
At $\lambda = 1$, i.e., the type-III case, $S$ is still negative but $|S|$ decreases. 
This indicates the non-monotonic behavior of $S$ upon increasing the tilting of the Dirac cones from the type-I side.
Then, the sign change of $S$ occurs for $\lambda = \lambda_{\rm c}$ with $\lambda_{\rm c} > 1$. 
Note that this behavior is also seen at finite temperatures, as shown in Fig.~\ref{fig:finite_L}(a).
The result can be understood as follows. 
From Eqs.~(\ref{eq:cond}) and (\ref{eq:L12_lowT}), 
one finds that $S$ is proportional to the $\mu$-derivative of $\sigma$ at $T= 0$.
Then, the sign of $S$, which is equal to that of $L_{12}$, is dictated by whether $\sigma$ at $\mu=0$ is a dip or a peak.
Clearly, types-I and III show a dip, while type-II shows a peak, which coincides with the resulting sign of $S$.

Next, we present the results of the power factor.
To estimate the low-temperature behavior, we again employ the Sommerfeld expansion to $L_{12}$.
In Fig.~\ref{fig:pf}(a)-(c), we show the $\mu$ dependence of PF.
As can clearly be seen, a large power factor is obtained for types I and III for small but finite $\mu$.
In particular, the power factor for the type-III Dirac system is the largest among the three types 
for $\Gamma = 0.02, 0.05$ eV with $|\mu| \lesssim 0.05$ eV, due to the subtle competition between $L_{12}$ and $L_{11}$.

\section{Summary \label{sec:summary}}
In this paper, we have investigated the electric and thermoelectric transport coefficients 
of the two-orbital square-lattice model in Eq.~(\ref{eq:Ham_sq}).
In this model, the type of Dirac cones can be tuned by a single parameter, $\lambda$,
and thus, the model serves as a minimal model for studying transport phenomena.

We have computed the electric conductivity, the Seebeck coefficient, the power factor, and the dimensionless figure of merit,
on the basis of the Kubo formula and the relaxation time approximation.  
We have found that the transport coefficients of the type-III case cannot be regarded as a simple limit of the type-I or type-II case. 
Actually, an optimal degree of tilting and chemical potential to obtain the largest thermoelectric responses
within the type-I regime exits; the type-III Dirac cone is not the optimal case. 
Furthermore, the chemical potential should not be right at the Dirac point. 
The best chemical potential for the large Seebeck coefficient will be near $\mu\sim 0.05$eV. 
As for the temperature dependence, the peaks appear at $T \sim \mu/2\Bolz$.
For the optimal case within our results, sizable transport coefficients are obtained; for example, the dimensionless figure of merit is 0.18.

To understand the physical origin of the above behaviors, we have also studied the low-temperature behaviors
by using the Mott formula.
We have found that
the sign of the Seebeck coefficient for the type-III case is the same as that for the type-I case.
This originates from the fact that the spectral conductivity shows a dip rather than a peak at $\epsilon = 0$.

Finally, we address the possible implications for real materials. 
The type-I Dirac cones with large tilting in quasi-two dimensions are realized in organic conductors 
such as $\alpha$-(BEDT-TTF)$_2$I$_3$~\cite{Katayama2006,Fukuyama2007,Goerbig2008,Kobayashi2008,Kobayashi2009} 
and $\alpha$-(BETS)$_2$I$_3$  [BETS is bis(ethylenedithio)tetraselenafulvalene]~\cite{Inokuchi1995,Tsumuraya2021,Kitou2021}.
For $\alpha$-(BEDT-TTF)$_2$I$_3$, the measurements of 
the Seebeck coefficients have indeed beeb reported~\cite{Konoike2013,Kitamura2014}.
Further interestingly, the degree of titling can be tuned by applying pressure~\cite{Kobayashi2004,Katayama2006,Kishigi2017}.
Therefore, these materials will be candidates for testing the tilting dependence of the thermoelectric 
transport coefficients.

\acknowledgements
We thank I. Tateishi and S. Ozaki for fruitful discussions and comments. 
T. M. thanks Y. Hatsugai for the collaboration in the prior work (Ref.~\cite{Mizoguchi2020_typeIII}).
This work is supported by JSPS KAKENHI, Grants
No.~JP18H01162, No.~JP18K03482, No.~JP19K03720, and No.~JP20K03802, and by the JST-Mirai Program, Grant No.~JPMJMI19A1.
T. M. is supported by JSPS KAKENHI, Grant No.~JP20K14371. 

\appendix
\section{Proof for $\sigma_{xx} = \sigma_{yy}$ \label{app:xy}}
In this appendix, we show a proof of the relation $\sigma_{xx} = \sigma_{yy}$ in the present model. 
Note that $C_4$ symmetry is broken in this model, thus the above relation is not obtained straightforwardly. 

Let $\alpha_{yy}(\epsilon)$ be the spectral conductivity for the $y$ direction;
that is, $\alpha_{yy}(\epsilon)$ is obtained by replacing $v_x(\bm{k})$ with $v_y(\bm{k})$ in Eq.~(\ref{eq:alpha}), as 
\begin{eqnarray}
\alpha_{yy} (\epsilon) &=& \frac{\hbar e^2}{2\pi A d_0} \sum_{\bm{k}}\mathrm{Tr} 
\{ G^{(R)}(\bm{k},\epsilon) v_y (\bm{k}) G^{(A)}(\bm{k},\epsilon) v_y (\bm{k}) \nonumber \\
&-&\mathrm{Re}  \left[G^{(R)}(\bm{k},\epsilon) v_y (\bm{k}) G^{(R)}(\bm{k},\epsilon) v_y (\bm{k})\right] \}.
\label{eq:alpha_bar}
\end{eqnarray}
In the following, we show that $\alpha_{xx}(\epsilon) = \alpha_{yy}(\epsilon)$ holds.
For simplicity of writing, we set $a_0 = 1$ in this appendix.

To begin with, we show that the spectral conductivity is an even function of $\epsilon$; that is, 
$\alpha_{xx}(\epsilon)  = \alpha_{xx} (- \epsilon)$ holds. 
To this aim, we first point out that $\mathcal{H}(k_x,k_y)$ satisfies
\begin{eqnarray}
\mathcal{H}(k_x+\pi,k_y+\pi) = - \mathcal{H}(k_x,k_y).
\end{eqnarray}
Therefore, we have 
\begin{eqnarray}
v_x (k_x+\pi,k_y+\pi) = - v_x (k_x,k_y), \label{eq:vx_change_1}
\end{eqnarray}
and 
\begin{eqnarray}
G^{(R)}(k_x + \pi ,k_y + \pi ,\epsilon) 
&=& [(\epsilon+i\Gamma) -  \mathcal{H}(k_x + \pi,k_y + \pi)]^{-1} \nonumber \\
&=&- [ (-\epsilon-i\Gamma) -\mathcal{H}(k_x,k_y) ]^{-1}\nonumber \\
&=& - G^{(A)}(k_x ,k_y,-\epsilon). \label{eq:gf_change_1}
\end{eqnarray}
Substituting Eqs.~(\ref{eq:vx_change_1}) and (\ref{eq:gf_change_1}) into Eq.~(\ref{eq:alpha})
and changing the variables as $k_x \rightarrow k_x-\pi $ and $k_y \rightarrow k_y-\pi$, 
we have 
\begin{widetext}
\begin{eqnarray}
\alpha_{xx}(\epsilon) &=& \sum_{k_x,k_y} \frac{\hbar e^2}{2\pi A d_0}\mathrm{Tr} 
\{ G^{(R)}(k_x,k_y, \epsilon)  v_x  (k_x,k_y) G^{(A)}(k_x,k_y, \epsilon)  v_x  (k_x,k_y)  \nonumber \\
&-&\mathrm{Re}  \left[G^{(R)}(k_x,k_y,\epsilon) v_x (k_x,k_y) G^{(R)}(k_x,k_y,\epsilon) v_x (k_x,k_y) \right] \} \nonumber \\
&=& \sum_{k_x,k_y} \frac{\hbar e^2}{2\pi A d_0} \mathrm{Tr} 
\{G^{(R)}(k_x + \pi,k_y+ \pi, \epsilon)  v_x  (k_x+ \pi,k_y+ \pi) G^{(A)}(k_x+ \pi,k_y+ \pi, \epsilon)  v_x  (k_x+ \pi,k_y+ \pi) \nonumber \\
&-&\mathrm{Re}  \left[G^{(R)}(k_x+ \pi,k_y+ \pi,\epsilon) v_x (k_x+ \pi,k_y+ \pi) G^{(R)}(k_x+ \pi,k_y+ \pi,\epsilon) v_x (k_x+ \pi,k_y+ \pi) \right] \}\nonumber \\
&=& \sum_{k_x,k_y} \frac{\hbar e^2}{2\pi A d_0} \mathrm{Tr} 
\{G^{(A)}(k_x,k_y, -\epsilon)  v_x  (k_x,k_y) G^{(R)}(k_x,k_y, -\epsilon)  v_x  (k_x,k_y)  \nonumber \\
&-&\mathrm{Re}  \left[G^{(A)}(k_x,k_y,-\epsilon) v_x (k_x,k_y) G^{(A)}(k_x,k_y,-\epsilon) v_x (k_x,k_y)  \right] \} \nonumber \\
 &=& \alpha_{xx}(-\epsilon). \label{eq:alpha_relation_1}
\end{eqnarray} 
\end{widetext}
Note that we have used $G^{(A)}(k_x,k_y,\epsilon) = [G^{(R)}(k_x,k_y, \epsilon)]^\ast$ and $v_x^\ast  (k_x,k_y)  = v_x  (k_x,k_y)$,
which lead to $\mathrm{Re}  \left[G^{(A)}(k_x,k_y,\epsilon) v_x (k_x,k_y) G^{(A)}(k_x,k_y,\epsilon) v_x (k_x,k_y)  \right] 
= \mathrm{Re}  \left[G^{(R)}(k_x,k_y,\epsilon) v_x (k_x,k_y) G^{(R)}(k_x,k_y,\epsilon) v_x (k_x,k_y)  \right]$.

Next, we show that $\alpha_{yy}(\epsilon) =\alpha_{xx}(-\epsilon)$.
To show this, we point out that $\mathcal{H}(k_x,k_y)$ satisfies 
\begin{eqnarray}
\mathcal{H}(k_x,k_y) = -\tau_x \mathcal{H}(k_y,k_x) \tau_x,
\end{eqnarray}
where $\tau_x$ is the $x$ component of the Pauli matrix.
Then, we have 
\begin{eqnarray}
v_y(k_x,k_y) &=&\frac{1}{ \hbar} \frac{\partial \mathcal{H}(k_x,k_y)}{\partial k_y} 
= -\frac{1}{ \hbar}\tau_x \frac{\partial  \mathcal{H}(k_y,k_x)}{\partial k_y} \tau_x \nonumber \\
&=& -\tau_x v_x (k_y,k_x) \tau_x, \label{eq:vx_and_vy}
\end{eqnarray}
and 
\begin{eqnarray}
G^{(R)}(k_x,k_y,\epsilon) 
&=& [(\epsilon+i\Gamma) -  \mathcal{H}(k_x,k_y)]^{-1} \nonumber \\
&=&- \{ \tau_x [(-\epsilon-i\Gamma) -\mathcal{H}(k_y,k_x)] \tau_x \}^{-1}\nonumber \\
&=& - \tau_x G^{(A)}(k_y,k_x,-\epsilon) \tau_x. \label{eq:gf_change_2}
\end{eqnarray}
Substituting Eqs.~(\ref{eq:vx_and_vy}) and (\ref{eq:gf_change_2}) into Eq.~(\ref{eq:alpha_bar})
and changing the variables as $k_x \rightarrow k_y $ and $k_y \rightarrow k_x$,
we have 
\begin{widetext}
\begin{eqnarray}
\alpha_{yy}(\epsilon) &=& \sum_{k_x,k_y} \frac{\hbar e^2}{2\pi A d_0}\mathrm{Tr} 
\{ G^{(R)}(k_x,k_y, \epsilon)  v_y (k_x,k_y) G^{(A)}(k_x,k_y, \epsilon)  v_y  (k_x,k_y)  \nonumber \\
&-&\mathrm{Re}  \left[G^{(R)}(k_x,k_y,\epsilon) v_y (k_x,k_y) G^{(R)}(k_x,k_y,\epsilon) v_y (k_x,k_y)  \right] \} \nonumber \\
&=& \sum_{k_x,k_y} \frac{\hbar e^2}{2\pi A d_0} \mathrm{Tr} 
\{ \tau_x G^{(A)}(k_y,k_x,-\epsilon)  v_x  (k_y,k_x) G^{(R)}(k_y,k_x,-\epsilon)  v_x  (k_y,k_x)\tau_x   \nonumber \\
&-&\mathrm{Re}  \left[\tau_x G^{(A)}(k_y,k_x,-\epsilon) v_x (k_y,k_x) G^{(A)}(k_y,k_x,- \epsilon) v_x (k_y,k_x) \tau_x \right] \} \nonumber \\
&=& \sum_{k_x,k_y} \frac{\hbar e^2}{2\pi A d_0} \mathrm{Tr} 
\{ \tau_x G^{(A)}(k_x,k_y,-\epsilon)  v_x  (k_x,k_y) G^{(R)}(k_x,k_y,-\epsilon)  v_x  (k_x,k_y)\tau_x   \nonumber \\
&-&\mathrm{Re}  \left[\tau_x G^{(A)}(k_x,k_y,-\epsilon) v_x (k_x,k_y) G^{(A)}(k_x,k_y,- \epsilon) v_x (k_x,k_y) \tau_x \right] \} \nonumber \\
 &=&\alpha_{xx} (-\epsilon). \label{eq:alpha_relation_2}
\end{eqnarray}
\end{widetext}
To obtain the final line of Eq.~(\ref{eq:alpha_relation_2}), we have used the fact that the trace is invariant under cyclic permutations.

Combining (\ref{eq:alpha_relation_1}) and (\ref{eq:alpha_relation_2}), we find $\alpha_{xx}(\epsilon) = \alpha_{yy}(\epsilon)$, which leads to $\sigma_{xx} = \sigma_{yy}$.

\section{Peak temperature of the Seebeck coefficient 
and the figure of merit for the type-I and type-III Dirac fermions \label{app:analytic}}
In this appendix, we elucidate the origin of the peak temperature of $S$, 
using the evaluation method proposed by Mahan and Sofo~\cite{Mahan1996}. 
Note that the same argument was presented in Ref.~\onlinecite{Hasdeo2019} for the conventional Dirac fermion system.
From Eqs. (\ref{eq:cond}), (\ref{eq:L12}) and (\ref{eq:L22}), we find
\begin{eqnarray}
L_{11} = \int_{-\infty}^{\infty} dw  \hspace{.5mm} g_0(w) \alpha_{xx} (w/\beta + \mu), \label{eq:L11_2}
\end{eqnarray}
\begin{eqnarray}
L_{12} =  -\frac{1}{e \beta} \int_{-\infty}^{\infty} dw \hspace{.5mm} g_1 (w)\alpha_{xx} (w/\beta + \mu),\label{eq:L12_2}
\end{eqnarray}
and 
\begin{eqnarray}
L_{22} =  \frac{1}{e^2 \beta^2} \int_{-\infty}^{\infty} dw \hspace{.5mm} g_2 (w)\alpha_{xx} (w/\beta + \mu),\label{eq:L12_2}
\end{eqnarray}
where $w:= \beta(\epsilon-\mu)$ and 
\begin{eqnarray}
g_n(w) = \frac{w^n e^w}{\left(e^w + 1\right)^2}.
\end{eqnarray}
Note that $g_n(w)$ is an odd (even) function of $w$ if $n$ is odd (even). 

Hereafter, we assume that $\mu$ is positive for simplicity. 
For analytical estimation of $L_{11}$, $L_{12}$ and $L_{22}$, 
we assume a simple analytic form of the spectral conductivity. 
Specifically, from the numerical results in Fig.~\ref{fig:cond}, 
for the type-I and type-III Dirac systems, the spectral conductivity 
around $\epsilon = 0$ can be approximated as 
\begin{eqnarray}
\alpha_{xx}(\epsilon) \sim \alpha_0+ \alpha_1 |\epsilon|, \label{eq:spectral_approximation}
\end{eqnarray}
where $\alpha_0$ and $ \alpha_1$ are coefficients.
Substituting Eq.~(\ref{eq:spectral_approximation}) into Eqs.~(\ref{eq:L11_2}) and (\ref{eq:L12_2}) 
and recalling the definition of $S$ in Eq.~(\ref{eq:Seebeck}), we find
\begin{eqnarray} 
S = -\frac{k_{\rm B}}{e} \cdot \frac{2\alpha_1 \mu h_1 (-w_0)+\alpha_1 k_{\rm B} T h_2(-w_0)}{\alpha_0 + 2 \alpha_1 k_{\rm B} Th_1(-w_0)  + \alpha_1 \mu h_0(-w_0)}. \nonumber \\
\end{eqnarray}
Here, $w_0:= - \mu/(k_{\rm B} T)$ and the functions $h_n(w)$ ($n=0,1,2$) are given as
\begin{subequations}
\begin{eqnarray}
h_0(w) = \int_{-w}^{w} dw^\prime \hspace{.5mm} g_0(w^\prime) = \tanh \frac{w}{2},
\end{eqnarray}
\begin{eqnarray}
h_1(w) = \int_{w}^{\infty} dw^\prime \hspace{.5mm} g_1(w^\prime) = \ln(1+e^w) - \frac{w e^w}{e^w + 1}, \nonumber \\
\end{eqnarray}
and 
\begin{eqnarray}
h_2 (w) &=& \int_{-w}^{w} dw^\prime \hspace{.5mm} g_2(w^\prime) \nonumber \\
&=& \frac{2 w^2 e^w}{e^w + 1} -4w \ln(1+e^w)-4\mathrm{Li}_2(-e^w) -\frac{\pi^2}{3}.\nonumber \\
\end{eqnarray}
\end{subequations}
Here, $Li_s(z)$ stands for the polylogarithm function of order $s$.
Note that the integration range of $w$ in Eqs.~(\ref{eq:L11_2}) and (\ref{eq:L12_2}) runs over $w\in  [-\infty,\infty]$, where 
the approximation of (\ref{eq:spectral_approximation}) breaks down. 
Nevertheless, using (\ref{eq:spectral_approximation}) is valid as long as 
$\mu$ is close to $0$, because $g_n(w)$ decays rapidly as $|w| \rightarrow \infty$. 
\begin{figure}[tb]
\begin{center}
\includegraphics[clip,width = 0.95\linewidth]{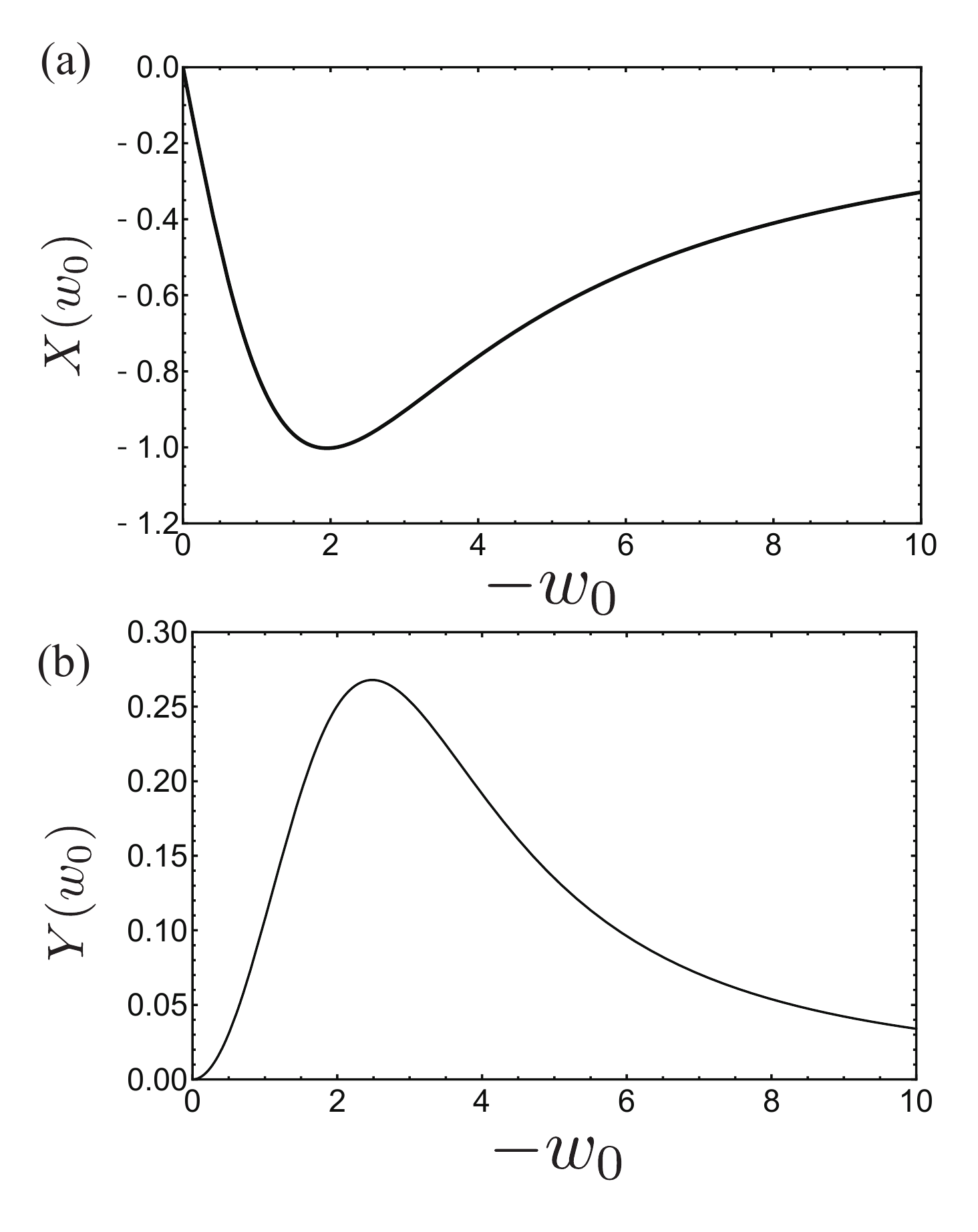}
\vspace{-10pt}
\caption{(a) $X(w_0)$ from Eq.~(\ref{eq:X}) 
and (b) $Y(w_0)$ from Eq.~(\ref{eq:Y}) 
as a function of $-w_0$.
Note that $w_0$ is negative when $\mu$ is positive.}
 \label{fig:analytic}
 \end{center}
 \vspace{-10pt}
\end{figure}

Further, Fig.~\ref{fig:cond} indicates that, in the clean limit 
(i.e, when $\Gamma$ is sufficiently small),
$\alpha_0$ in the spectral conductivity becomes less dominant.
Hence, we set $\alpha_0\rightarrow 0$ for simplicity.
By doing so, we have 
\begin{eqnarray} 
S &\sim&\frac{k_{\rm B}}{e} \cdot X(w_0), \nonumber \\
X(w_0) &=& - \frac{ h_2(-w_0)- 2 w_0 h_1 (-w_0)}
{2 h_1(-w_0) - w_0 h_0(- w_0)}, \nonumber \\ \label{eq:X}
\end{eqnarray}
which does not depend on $\alpha_1$.
Equation (\ref{eq:X}) indicates that the temperature and 
chemical potential dependence of $S$ is determined by the single variable $w_0 = -\mu/(k_{\rm B} T)$. 

Figure~\ref{fig:analytic}(a) shows the function $X(w_0)$
for positive $\mu$ (i.e., negative $w_0$).
We see that the peak of $X(w_0)$ is indeed at $-w_0 \sim 2$, 
i.e., $T \sim \mu/(2 k_{\rm B})$, 
which coincides with the numerical result shown in Fig.~\ref{fig:finite}(d). 
We also see that the peak height of $|X(w_0)|$ is almost 1,
meaning that the maximal $|S|$ within this approximation is $k_{\rm B}/e \sim 86$ $\mu$V/K.
In actual numerical numerical calculation [Fig.~\ref{fig:finite}(d)], the peak height is smaller than the above value and it also
depends on $\mu$, which might be because $\alpha_0$ is non-negligible.  

The estimation of $ZT$ can be performed in the same way.
Again neglecting $\alpha_0$, we have
\begin{widetext}
\begin{eqnarray}
ZT = Y(w_0), \hspace{.5mm}
Y(w_0) = \left[\frac{[2h_1(-w_0) -w_0 h_0(-w_0)][2h_3(-w_0)-w_0 h_2(-w_0)]}{[h_2(-w_0)-2 w_0 h_1(-w_0)]^2} - 1\right]^{-1}, \label{eq:Y}
\end{eqnarray}
where 
\begin{eqnarray}
h_3(w) &=& \int_w^{\infty} dw^\prime \hspace{.5mm} g_3(w^\prime) 
 = w^2 \left[ 3\ln \left(1 +e^w\right) -\frac{w e^w}{e^w + 1} \right] + 6w \mathrm{Li}_2(-e^w) -6 \mathrm{Li}_3(-e^w). \nonumber \\
 \end{eqnarray}
 \end{widetext}
Figure~\ref{fig:analytic}(b) shows the function $Y(w_0)$.
We see that the peak of $Y(w_0)$ is $-w_0 \sim 2.5$.
Thus, the peak temperature of $ZT$ is $T \sim \mu/(2.5 k_{\rm B})$, which is slightly smaller than that for $S$.
We also see that the maximum of $ZT$ is about 0.27.
This value is greater than the optimal $ZT$ obtained in the numerical calculation, which might again be due to the effect of $\alpha_0$.

\section{Role of Dirac points in the Seebeck coefficient for the type-II case \label{app:typeII}}
\begin{figure*}[tb]
\begin{center}
\includegraphics[clip,width = 0.95\linewidth]{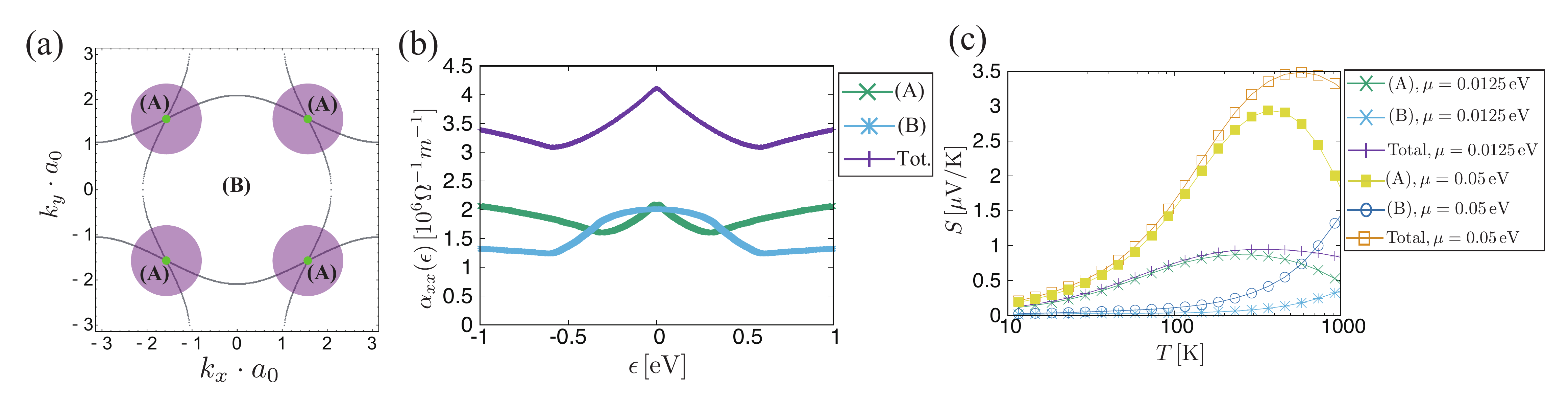}
\vspace{-10pt}
\caption{(a) Schematic figure of the division of the Brillouin zone into the two regions, (A) and (B). 
The region (A) is composed of four circles whose radii are $\frac{\pi}{4a_0}$ and centers are the Dirac points. 
The gray lines are the Fermi surface for $\mu = 0$ eV. 
(b) $\alpha_{xx}^{\rm(A)} (\epsilon)$, $\alpha_{xx}^{\rm(B)} (\epsilon)$, and $\alpha_{xx}(\epsilon)$ as functions of $\epsilon$.
(c) Temperature dependence of $S^{\rm (A)}$, $S^{\rm (B)}$, and $S$. 
For (b) and (c), we set $\Gamma = 0.02$ eV.}
 \label{fig:typeII}
 \end{center}
 \vspace{-10pt}
\end{figure*}
In this appendix, we clarify how the Dirac points contribute to the Seebeck coefficient.
For the type-II case, the Fermi surface extends far away from the Dirac points [Fig.~\ref{fig:band}(g)];
thus, it is worth investigating the contribution from the region near the Dirac points and those from the rest separately. 

To do this, we first divide the $\bm{k}$ space into two regions: 
One is the vicinity of the Dirac points, which we call (A), and the other is the remainder, which we call (B) [see Fig.~\ref{fig:typeII}(a)].
Then, the spectral conductivity in Eq.~(\ref{eq:alpha}) 
can be divided into two contributions by restricting the summation over $\bm{k}$ to either (A) or (B).
We call these contributions $\alpha_{xx}^{\rm (A)} (\epsilon)$ and $\alpha_{xx}^{\rm(B)} (\epsilon)$, respectively. 
In Fig.~\ref{fig:typeII}(b), we plot $\alpha_{xx}^{\rm(A)} (\epsilon)$ and $\alpha_{xx}^{\rm(B)} (\epsilon)$.
We see that these two contributions are comparable near $\epsilon =0$ eV.
Therefore, as far as the electric conductivity is concerned, the Dirac points do not have special importance. 

As for the Seebeck coefficient, by substituting $\alpha_{xx}^{\rm(A)} (\epsilon)$ and $\alpha_{xx}^{\rm(B)} (\epsilon)$ into Eq.~(\ref{eq:L12}),
we obtain $L_{12}^{\rm(A)}$ and $L_{12}^{\rm(B)}$, respectively. 
Using these, we define 
\begin{eqnarray}
S^{\rm (A)/(B)} = \frac{1}{T} \frac{L^{\rm (A)/(B)}_{12}}{L_{11}}.
\end{eqnarray}
In Fig.~\ref{fig:typeII}(c), we plot $S^{(A)}$, $S^{(B)}$ and $S$ as functions of $T$ for $\mu= 0.0125$ eV and $0.05$ eV. 
We see that the large contribution to $S$ comes from region (A) in both cases, 
which implies that the Dirac points play an important role in the thermoelectric transport in this system. 

\bibliographystyle{apsrev4-2}
\bibliography{Dirac}
\end{document}